# The environmental impact of ICT in the era of data and artificial intelligence

François Rottenberg, Thomas Feys, Liesbet Van der Perre.

07/01/26

**François Rottenberg** is an Assistant Professor at the Department of Electrical Engineering (ESAT) at KU Leuven, in the Dramco research group. His main research activities are in developing power consumption models, and designing signal processing and machine learning algorithms to enable sustainable data communications.

**Thomas Feys** is a PhD student at the ESAT-Dramco research group at KU Leuven. He previously obtained the M.Sc. degree in electronics and ICT engineering technology and the advanced M.Sc. degree in Artificial Intelligence from KU Leuven. His main research interests are machine learning for energy efficient wireless communications.

**Liesbet Van der Perre** is a Full Professor at the ESAT-Dramco research group at KU Leuven and a guest Professor at ULund, Sweden. Her main research interests are in wireless communication and embedded connected systems, with a focus on sustainable solutions.

**Abstract**: The technology industry promotes artificial intelligence (AI) as a key enabler to solve a vast number of problems, including the environmental crisis. However, when looking at the emissions of datacenters from worldwide service providers, we observe a rapid increase aligned with the advent of AI. Some actors justify it by claiming that the increase of emissions for digital infrastructures is acceptable as it could help the decarbonization of other sectors, e.g., videoconference tools instead of taking the plane for a meeting abroad, or using AI to optimize and reduce energy consumption. With such conflicting claims and ambitions, it is unclear how the net environmental impact of AI could be quantified. The answer is prone to uncertainty for different reasons, among others: lack of transparency, interference with market expectations, lack of standardized methodology for quantifying direct and indirect impact, and the quick evolutions of models and their requirements.

This report provides answers and clarifications to these different elements. Firstly, we consider the direct environmental impact of AI from a top-down approach, starting from general information and communication technologies (ICT) and then zooming in on data centers and the different phases of AI development and deployment. Secondly, a framework is introduced on how to assess both the direct and indirect impact of AI. Finally, we finish with good practices and what we can do to reduce AI impact.

**Disclaimer**: The authors are happy to share that this document was written solely based on their own generative investigation.

**Acknowledgement**: The authors thank their KU Leuven colleagues and experts for their relevant review and feedback, which they integrated in the document.

# 1. Introduction: conflicting claims and ambitions

The chief sustainability officer from Microsoft stated on LinkedIn: "AI and the successful energy transition go hand in hand. We will not see a large-scale shift to carbon-free energy without the significant advancements AI promises to deliver." Along this line, the tech giants generally sell artificial intelligence (AI) as one key solution to solve a number of problems, including the environmental crisis. On the other hand, when looking at the emissions of Microsoft and Google (see Figure 1), we observe an increase at a fast pace, which did not particularly slow down with the advent of AI in 2022: a 54% increase for Google in 2024 since 2019 [1] and a 23.4% for Microsoft in 2024 since 2020 [2]. This is in conflict with their climate ambitions to reach carbon neutrality by 2030.[1]

Google states [1]: "As we further integrate AI into our products, reducing emissions may be challenging due to increasing energy demands from the greater intensity of AI compute, and the emissions associated with the expected increases in our technical infrastructure investment". Indeed, despite its virtual nature, AI has a clear physical footprint related to its required digital infrastructure: resource extraction required to produce the hardware for training and running the models, storing the data, generating electricity, data centers, cooling… At the same time, some actors claim that the increase of emissions for digital infrastructures is acceptable as it could help to decarbonize other sectors and allow net reductions, e.g., using videoconference tools instead of taking the plane for a meeting abroad, or using AI to optimize energy consumption by using smart warming. Along this line, the European Commission sees it as a key ingredient to reach climate neutrality in 2050 [4]: « Europe must leverage the potential of the digital transformation, which is a key enabler for reaching the Green Deal objectives. »

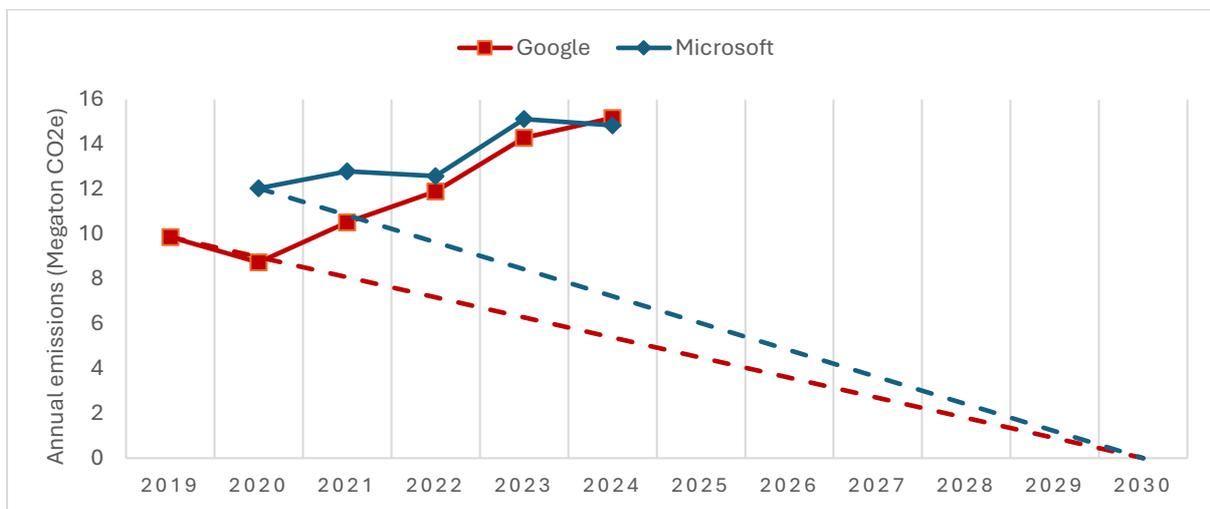

Figure 1 - Microsoft and Google's annual emissions in millions of tons of CO2 equivalent (continuous line) compared to their trajectories to reach their net-zero commitment by 2030 (dashed lines) (data source: [1], [2]).

Conflicting claims and ambitions – that is clear. However, how can we get a clear idea on the net environmental impact of AI? The answer is prone to uncertainty for different reasons:

---

[1]Studies have criticized the emissions reporting of major tech companies. Data center emissions could be much higher than reported, according to The Guardian [3]. One key critic is the use of market-based emissions allowing companies to buy renewable energy certificates and to match the related emissions (instead of using location-based emissions, related to the actual emissions from the area where and the time when the processing occurs).

1. AI has a virtual nature but well with a physical footprint, that is not directly visible and this virtual-physical relation is not straightforward to identify. For instance, it requires some investigation to relate a ChatGPT query of a user in Europe to the server in the US computing the result of the request.
2. There is a lack of transparency with few industrial data available and many of the estimates requiring some degree of reverse engineering.
3. There is also a lack of a standardized impact assessment methodology.
4. Technology is evolving very quickly, energy efficiency is improving, and energy consumption depends on many factors. As an example, the same ChatGPT query could have a different impact depending on its date. Moreover, which ChatGPT model version are/were you using? GPT-4o, GPT-3.5, GPT-4o mini, GPT-4.5, GPT-4 Turbo…? On top of that, for GPT-5, the query is routed to a different model depending on the complexity of the task. This makes the effective model size invisible.
5. Even when getting access to data, it is not easy to get a grasp on what it actually means in physical terms. What does the 500 tons CO2e required to train GPT-3 physically represent? Is it large or small compared to more tangible daily life equivalents?
6. Finally, how to make the net balance between positive (reduction) and negative impacts on the environment?

This document provides some answers and clarifications to these different elements. Section 2. considers the <u>direct</u> environmental impact of AI from a top-down approach, starting from general information and communication technologies (ICT) and then zooming in on data centers and then the different phases of AI. In Section 3., a framework is introduced on how to assess both the direct and the <u>indirect</u> impact of AI. Finally, we will finish Section 4 with <u>good practices</u> and what we can do as a user to reduce AI impact.

In other words, how to use AI technologies responsibly? And how to use AI for the good cause?

A few notes before starting. This document provides an informative summary for a general public. Sections 2 and 3, on the direct and indirect environmental impacts of AI, have a strong scientific basis as they review the relevant literature. On the other hand, Section 4 has a looser scientific foundation as it looks at a more practical daily approach. As a last note, the scope of this document only (mostly) covers environmental aspects of AI. This is of course a limited vision and it can create a sharp dichotomy between applications that are environmental-friendly or not. There are more aspects to consider. In Section 4, we advocate for a holistic sustainable vision of AI under its three fundamental dimensions: environment, society, and economy. For a given AI application, we should look at the cost-benefit balance along the three dimensions together. A use case, e.g., digital entertainment, could have a certain negative environmental impact, but which is outweighed by its social and/or economic benefits.

## 2. What is the direct environmental impact of ICT and AI?

In this section, we consider the direct environmental impact of raw material extraction, production, use, and disposal, which can be assessed through a life cycle assessment (LCA). The environmental impact is defined by the European Commission by looking at 16 indicators [5], including for instance greenhouse gas (GHG) emissions, water use, particulate matters or acidification. While all phases and impact categories of the LCA are important to consider, emphasis has been mostly on the GHG emissions and the production-use phase, mostly due to lack of transparency on the full life cycle of a product. As a result, most studies do not provide a complete picture of the impact. As an additional note, greenhouse gas emissions (GHG) are a commonly used indicator to assess global warming. These gases include carbon dioxide, methane, or nitrous oxide. For convenience, they are converted as a function of their global warming potential and lifetime into a single unit, the ton of CO2-equivalent (tCO2e).

### A. Information and communication technologies (ICT)

In 2021, a study review estimated that global emissions from ICT could reach 2.1%-3.9% of global greenhouse gas emissions [6], with 30% coming from embodied emissions (related to extraction of raw materials, manufacturing process and transport to the business/user) and 70% from the use phase. As a comparison, the International Energy Agency estimated in 2023 that the aviation sector accounted for 2.5% of energy-related $CO_2$ emissions [7]. In other words, both sectors have emissions in the same order (ICT could even be larger depending on the estimate). The ICT footprint is generally further divided into three main components: data centers, networks and user devices. In the following, we zoom in on the first one, most relevant for AI.

### B. Data centers

#### Electricity consumption

According to the International Energy Agency (IEA), data centers accounted for about 1.5% of the world's electricity consumption in 2024 [8] reaching about 415 TWh in 2024. In Figure 2, we can observe a fast acceleration in 2017 (+12%/year growth), especially in the USA which can be attributed to the advent of cloud computing, online media consumption, social media platforms and the rise of AI.

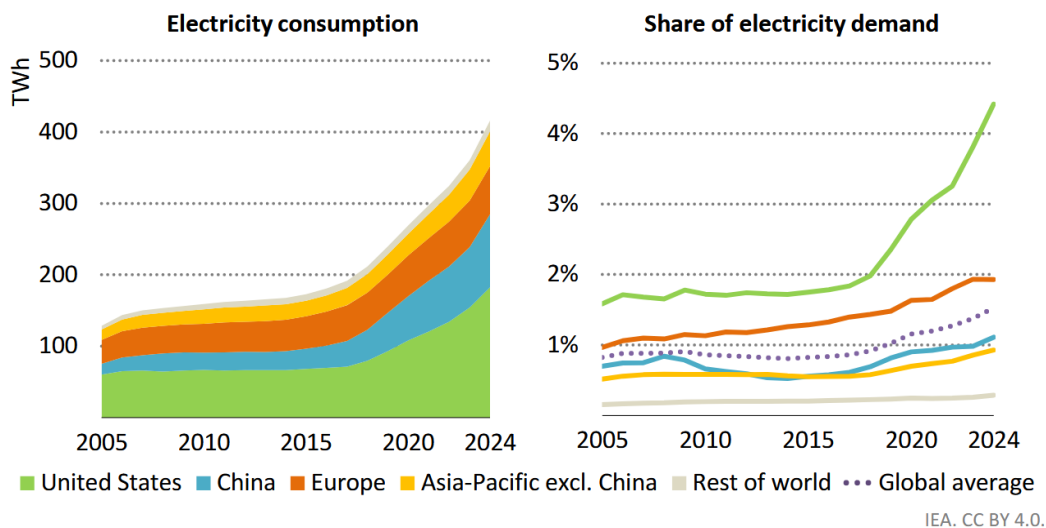

Figure 2 - Data centers' electricity consumption and share of demand across regions of the world and over time [8].

The same report considers different scenarios for the evolution of data center electricity demand. In the base case (current regulatory conditions and industry projections), the data center electricity consumption will more than double to around 1000 TWh by 2030, mostly due to AI and representing a 3% share of total electricity consumption or more than Japan's total electricity consumption today. In 2030, the USA would consume more electricity for data centers than for the industrial production of aluminum, steel, cement, chemicals and all other energy-intensive goods combined. In Belgium, the fast electrification of industries is slowed down as the grid is strongly congested . The data centers are/will be part of the cause of the problem [9], [10].

> **What share of the data center electricity demand comes from AI?**
>
> AI is only one of the workloads performed by data centers. Unfortunately, estimates in the literature vary widely and the AI share of electricity demand is difficult to obtain. The IEA has summarized data from different studies in its report (Figures 2.5 in [8]) showing a large variability, but well with an increasing trend (slower or faster depending on the study). They explain this variability by several reasons. Firstly, a clear distinction between AI-related and non-AI-related tasks is becoming more challenging. Secondly, there is a lack of comprehensive data with sufficient information on the energy consumption related to each workload. Thirdly, there are often differences in AI definitions.

## Carbon footprint

We need to know the carbon intensity of electricity production to convert electricity consumption to a carbon footprint. Due to the significant and fast increasing demand from data centers, about half of the electricity (see Figure 3) is expected in the near term to come from gas and coal energy sources [8], having a large carbon intensity. A data center also has to work 24 hours a day and cannot easily delay its workload when wind and/or solar low-carbon energy is not available. To counteract this trend, tech companies have planned to create new nuclear plants, producing electricity at a low carbon intensity. However, this will take years to be realized and only cover one part of the demand. According to [8], emissions are expected to approximately double by 2030 reaching 320 Mt CO2e, about 1% of total CO2 emissions. This hence counteracts climate objectives, when considered as a stand-alone contribution.

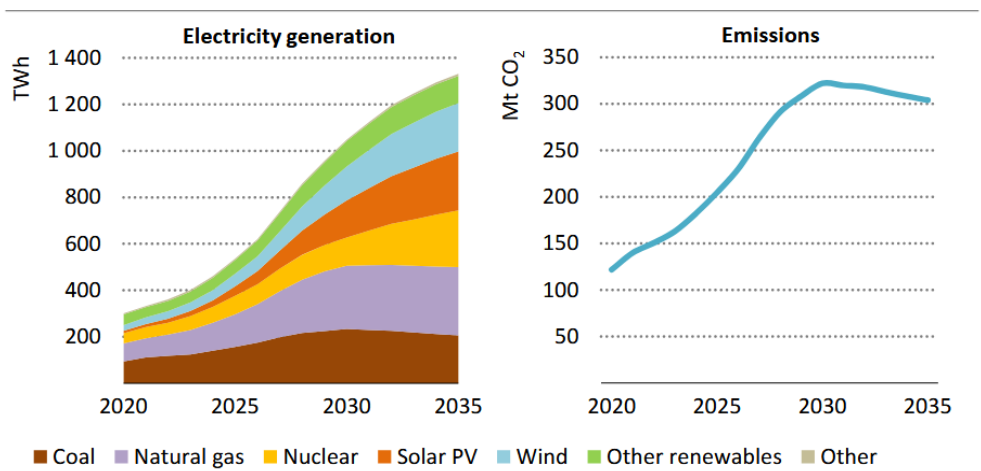

Figure 3 – Global electricity generation for data centers and associated emissions for 2020-2035 [8].

## Water usage

Data centers require water directly onsite for cooling as well as indirectly for electricity generation and (a.o. semiconductor) manufacturing. The water use can vary significantly depending on cooling technology, local climate and electricity supply. The quality (drinkable versus river water) and the abundance/scarcity at a given location can also vary a lot. We can distinguish between water withdrawals and consumption. Withdrawal is the total amount of water used by data centers. Consumed water represents the portion not returned which is lost due to evaporation for instance. As shown in Figure 4 [8], the IEA estimates global water widrawals around 5500 billions litres and consumption of around 560 billion litres in the year 2023. This consumption could more than double in 2030 reaching 1200 billion liters if current trends persist.

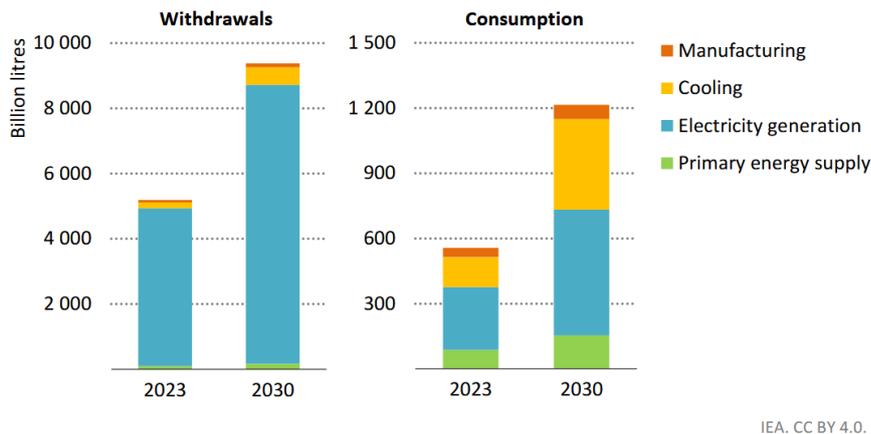

Figure 4 – Water consumption in billion liters by data centers in 2023 and projection in 2030 [8].

## Global versus local impact

Besides these global impacts, it is important to understand the local impacts of data centers. According to the IEA, the average consumption of one large data center can be equivalent to that of 100,000 households (100 MW). The largest under construction would consume the equivalent of 2 million households. However, data centers are much more geographically concentrated. Almost half of data center capacity in the USA is clustered in five regions. In Ireland (see Figure 5), about 20% of the electricity is consumed by data centers [11]. This creates competition between citizens, public services and data centers. The very localized energy demand of data centers also creates tension and bottlenecks on the network electricity grid. Similarly for water consumption, the very localized consumption competes with basic housing, agricultural or industrial activities, especially in case of drought.

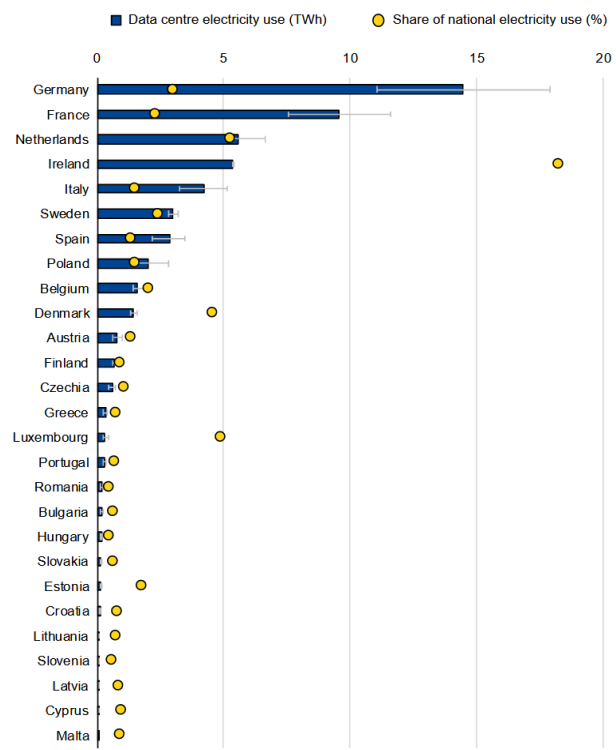

Figure 5 - Estimated data center energy use by country (2022) [11].

We note that the location of data centers has become a strategic matter, whereby ecological, economical, and autonomy in a geopolitical context are criteria that may get a higher priority in decisions. For example, building data centers where natural resources, in particular energy, are available, has been advocated as a solution to the ecological footprint, such as presented in Östersund, Sweden. However, one should be aware that the investment in energy-hungry installations in the North risks to worsen the energy shortage in the South of Sweden.

From a perspective of usage of computing and storage resources, central operation of very large datacenters is most efficient. However, deploying more distributed smaller datacenters, enhancing local autonomy[2], may bring a benefit of raising awareness among end-users of the energy and water consumption of the storage of data and AI. Awareness could further contribute to a more responsible attitude and slow down the trend of turning ever more frequently to AI.

## C. Different phases of AI and machine learning

Artificial intelligence is broadly defined as the ability of computer programs to perform actions that typically require human intelligence [12]. Given the broad scope of this definition, AI is a very wide field, including perception, learning, reasoning, problem-solving, etc. The current rise of generative AI, i.e., artificial intelligence that generates text, images, videos and so on, is largely based on machine learning. Machine learning can be seen as a subfield of AI where algorithms learn to perform a certain task from data. This paradigm is in stark contrast to classical algorithms which must be programmed. Imagine that a classical algorithm is a recipe to make a tasty dish, where each step has to be explained in detail to the computer, whereas a machine learning algorithm learns how to do this from data, i.e., from observing many tasty dishes. In sum, a classical algorithm has to be developed and programmed by a human programmer, whereas a machine learning algorithm is trained to perform a certain task by feeding it a large amount of data. This machine learning paradigm is the driving force behind popular large language models (LLMs) such as ChatGPT, Claude, Gemini, etc.

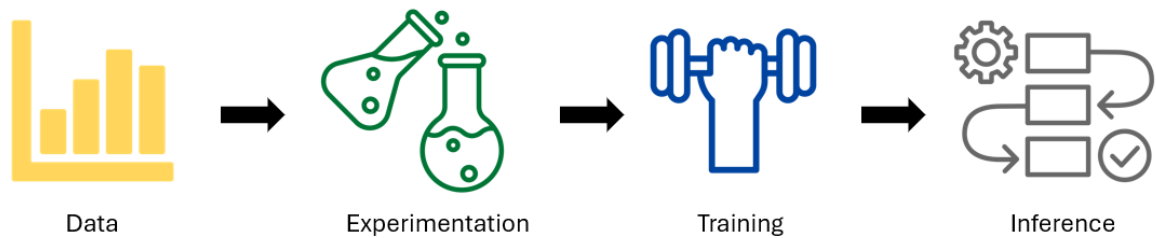

Figure 6 - Different phases of machine learning model development and deployment.

The four phases of this machine learning paradigm are illustrated in Figure 6: data collection, experimentation, training and deployment (inference). We here provide estimates for the impact of each phase. Again, this is not straightforward given the lack of transparency of available data. Industries do not publish much data and where there is, it is not always peer-reviewed or standardized so that making comparison is difficult. As an alternative, much work has performed some sort of reverse-engineering to evaluate the impact of each phase.

In an article from Facebook AI from 2022 and as shown in Figure 7, the authors report an average power breakdown of 10%-20%-70% for the three key phases experimentation-training-inference over a vast area of applications (not only LLM but also recommendation models for instance) [13]. The authors also stress that this breakdown is heavily dependent on the applications.

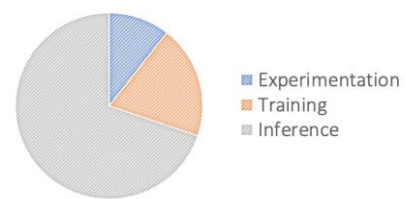

Figure 7 – Share of electricity consumption of each phase at Facebook datacenters [13].

---

[2]One could hence argue that investments in datacenters e.g. as recently announced in Belgium, should be welcomed. However, owned by foreign big tech companies, there's a risk that these do not enhance the country's autonomy.

Let us now look more specifically at each phase.

## Data

Collecting, processing and storing data to train machine learning has a significant cost. While getting exact numbers on this impact is hard, the trend is clear. There is an exponential increase in the dataset sizes required to train the models. Looking at large language models, Figure 8 shows the evolution of the number of tokens that were required to train some of the popular models [14]. Roughly speaking, a token corresponds to a word or a part of a word. On average, a typical book would contain 100 000 tokens and a classical tokenizer would need about 1.5 byte to encode one token. If a typical library shelf contains 100 books, it took about 2800 library shelves (42 Gigabytes) to train GPT2 in 2018, 30 000 library shelves (450 Gigabytes) to train GPT3 in 2020 and 1 000 000 library shelves (15 Terabytes) to train Llama 3 in 2024, a model from Meta.

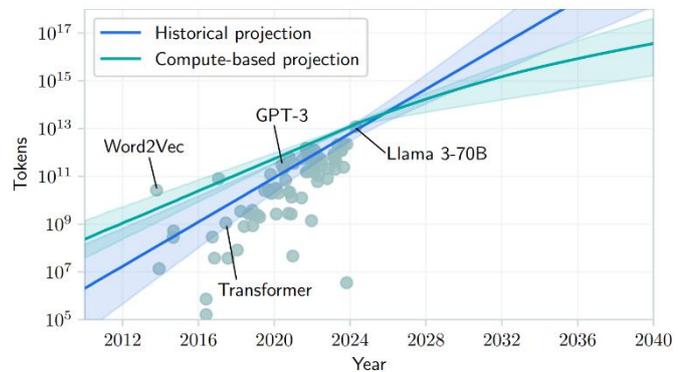

Figure 8 – Evolution of dataset size (expressed in tokens) across time for popular large language models [14].

## Experimentation

During experimentation, researchers explore the type and architecture of the machine learning model, the design of their hyperparameters, the performance of the algorithms, the training techniques… This model exploration is computationally intensive and a large number of models/ideas are explored in parallel. In this experimentation phase, many models are trained or retrained, up to 5000 times, which causes a significant energy consumption [15]. However, to keep it scalable, the training occurs only based on a subset of the dataset and using only a part of computing resources. At the end of the experimentation phase, the model is then trained on the full dataset and using all computing resources, which can take weeks.

## Training

Most research on estimating the carbon footprint of training a machine learning model uses some sort of reverse engineering due to the absence of data on the exact power consumption of the graphics processing units (GPUs)[3] [16]. This is typically calculated by multiplying the number of GPU hours by the power consumption of the GPU, the power usage effectiveness (PUE)[4] of data centers and finally the carbon intensity of the network grid. Figure 9 from [16] compares the training footprint of several large language models, collected from different studies. Again, since no standardized methodologies are established, one should be careful when comparing them.

---

[3]A GPU is a specialized electronic circuit, initially designed for accelerating graphics and video rendering, before being more widely applied in other fields like AI where they excel in data processing and computing tasks.
[4]The PUE is defined as the ratio of the total energy used by a data center facility divided by the energy consumed by the computing equipment. A lower PUE means a better efficiency and less energy wasted on overhead (cooling, power distribution…).

| Model name | Number of parameters | Datacenter PUE | Carbon intensity of grid used | Power consumption | $CO_2eq$ emissions | $CO_2eq$ emissions × PUE |
|---|---|---|---|---|---|---|
| GPT-3 | 175B | 1.1 | 429 g$CO_2$eq/kWh | 1,287 MWh | *502 tonnes* | 552 tonnes |
| Gopher | 280B | 1.08 | 330 g$CO_2$eq/kWh | *1,066 MWh* | *352 tonnes* | 380 tonnes |
| OPT | 175B | *1.09* | *231g$CO_2$eq/kWh* | *324 MWh* | 70 tonnes | *76.3 tonnes* |
| BLOOM | 176B | 1.2 | 57 g$CO_2$eq/kWh | 433 MWh | 25 tonnes | 30 tonnes |

Figure 9 – Comparison of carbon footprint of training a number of large language models [16].

The work in [17] (Figure 10) estimates the GPT-4 training emissions to 3088tCO2e (corresponding to 7200 MWh). In equivalent terms of long-haul flights considering a carbon intensity of 148g of CO2e per passenger kilometer [18], this corresponds to the emissions of about 1900 passengers flying between Beijing and New York City (11,000 km). When comparing to the average global human carbon footprint of 4.7 tons/year [19], this corresponds to emissions of an average human during 617 years.

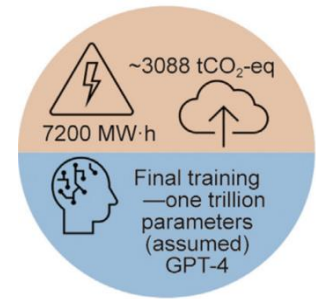

Figure 10 - GPT-4 training emissions [17].

## Inference

**Explosion of the number of users and queries.** A key aspect with the advent of generative AI is that the inference phase becomes the most resource-hungry (more than experimentation and training) due to the explosion of the usage. This can be surprising at first as a single query might not seem to consume much. In line, Sam Altman, the chief executive officer (CEO) of OpenAI, which launched ChatGPT, announced in June 2025 on his blog that "the average query uses about 0.34 watt-hours, about what an oven would use in a little over one second, or a high-efficiency lightbulb would use in a couple of minutes. It also uses about 0.000085 gallons of water; roughly one fifteenth of a teaspoon." Similarly, Google released a study stating: "we find the median Gemini Apps text prompt consumes 0.24 Wh of energy—a figure substantially lower than many public estimates. We also show that Google's software efficiency efforts and clean energy procurement have driven a 33x reduction in energy consumption and a 44x reduction in carbon footprint for the median Gemini Apps text prompt over one year." [20]. Unfortunately, the transparency of these (not peer-reviewed) studies is limited, especially the blog post with few details on the scope and the methodology. Some peer-reviewed studies found similar numbers though, such as the work of [21], which experimentally studied unoptimized Llama models and reported numbers of about 3 Joules per token. Assuming a typical number of output tokens equal to 500 (roughly equivalent to 400 words or one page of text), this gives 1500 Joules/query or 0.42 Wh/query. The online website Epoch AI provides some further information on the subject, discusses contradicting sources and how it compares to daily electricity equivalents [22]. Using reverse engineering and a few assumptions, it comes down to the similar figure of 0.3 Wh/query for the GPT-4o model.

These relatively low numbers may seem to imply a low inference impact, which is even decreasing over time given the efficiency gains. However, as seen in Figures 1-3, we should not forget that the global impact of AI and datacenters is well on the rise. Indeed, while the average energy consumption for a single query might not be so high, it is really the massive adoption of general public AI tools (see Figure 11) that makes it so power hungry and explains why the inference phase outweighs the training phase. AI is being progressively integrated into our daily life, often even in invisible ways. For instance, using an app or visiting a website could trigger the use of an AI tool for making recommendations, without specifically asking for it.

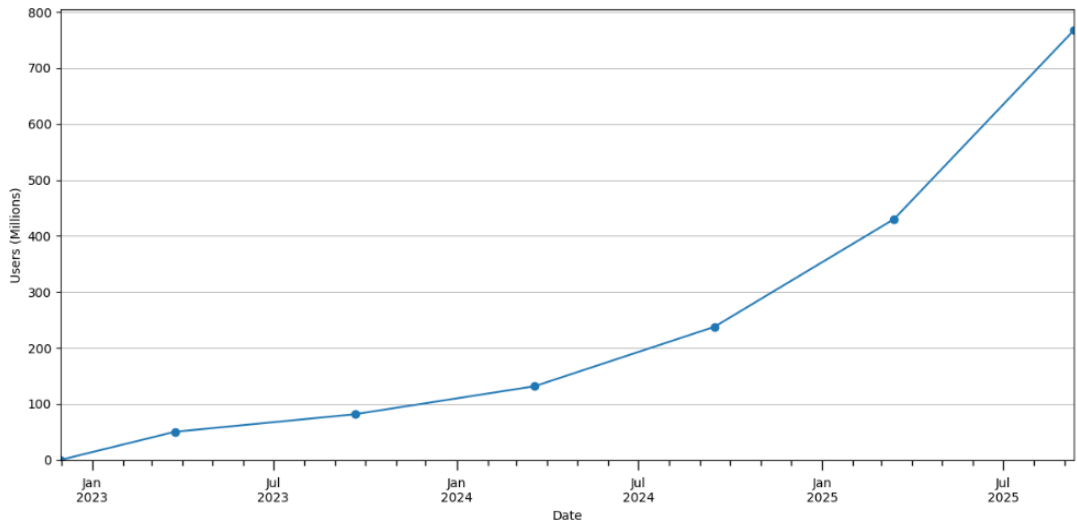

Figure 11 – Number of ChatGPT users per week on consumer plans (Free, Plus, Pro) [23].

To get a quantitative rough estimate, we can rely on usage data from [23]. By July 2025, more than 2.5 billion individual queries were sent per day. If we take the number of 0.3 Wh/query, this translates into 750 MWh of electricity per day. As compared to the 7200 MWh reported in [17] for the training of GPT-4, this would imply that the power consumption related to 10 days of inference is larger than the amount required to train the full model. While these numbers provide a rough approximation, the key takeaway message is that the inference cost becomes quickly larger than the training cost.

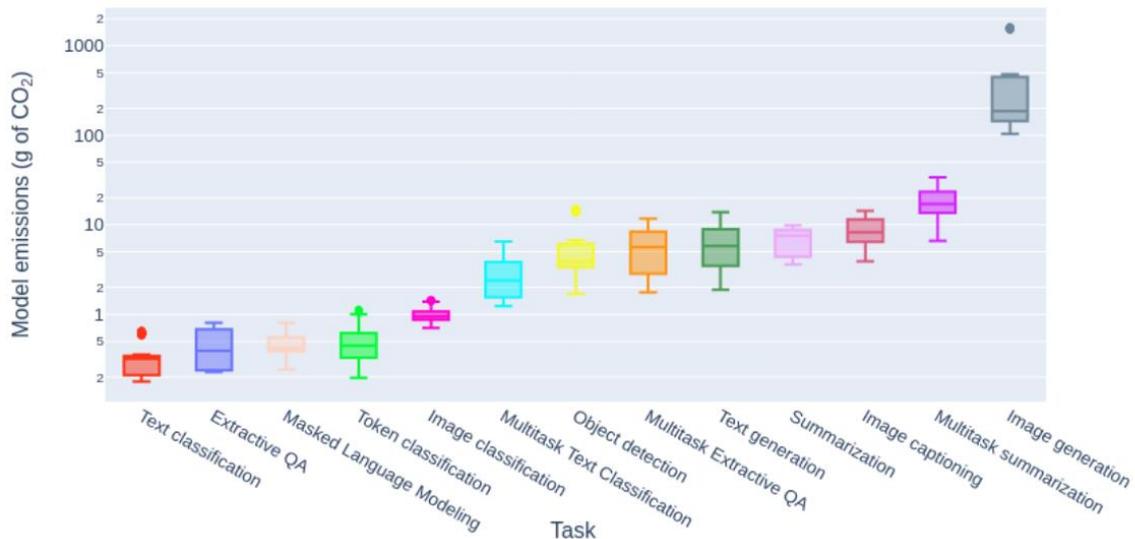

Figure 12 – Model emissions as a function of the task [24].

**Nature of the task.** Looking closer, the impact during inference is highly variable and depends on different factors. As mentioned, there is a lack of clear data to make firm claims due to the lack of transparency. Still, researchers have shown clear trends. As illustrated in Figure 12 [24], the emissions during inference are highly depending on the nature of the task, with summaries, image and text generation, being up to 10-100 times more power hungry than basic Q&A or classification tasks (e.g. spam detection). We should also be careful about making a general statement that image generation takes more energy than text generation. It might be surprising, but long text generation typically consumes more than image generation. Quoting interviews with experts reported in [25], "Large [text] models have a lot of parameters", "Even though they are generating text, they are doing a lot of work. " On the other hand, image generation often works with fewer parameters.

The IEA report [8] mentions that the generation of a 6 second video with 8 frames per second consumes as much energy as two laptop full charges, about 115 Wh. So, it seems that video generation requires a significant amount of energy, especially as compared to other AI tasks [25].

**Model type and size**. Secondly, besides the nature of the task, model type and size (number of parameters) have a strong influence. As an example, let us compare the energy consumption of different models using the Ecologits calculator (consulted in Dec. 2025): ChatGPT 4 (1760 billion parameters), ChatGPT 4 Turbo (880 billion parameters) and ChatGPT-4o (440 billion parameters). Note that the number of parameters are estimated as the official numbers were not disclosed. For a report of 5 pages (5000 output tokens) with the world average mix, the models require 100 Wh, 29.7 Wh and 9.61 Wh, respectively. Reducing by 2 the number of parameters reduces consumption by a factor of 3 to 4. It is important to note that these models do not only differ by their sizes but also in architecture, sparsity and the hardware they rely on. As shown in Figure 13 [26], the model size increases fast: 10 times more parameters for GPT-4 versus GPT-3, about three years before.

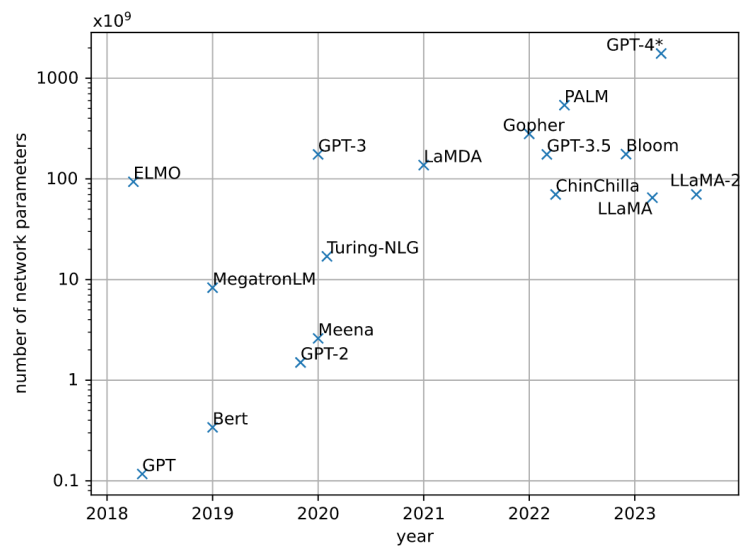

Figure 13 – Evolution of the number of parameters of popular models over the years [26]. *The number of parameters of GPT-4 is estimated as it was not officially confirmed.

Model distillation and routing are advanced AI techniques that are used to improve the trade-off between model size, performance and efficiency. Model distillation consists in "compressing" the knowledge from a large, computationally expensive model to a smaller, less complex and more energy-efficient one, while keeping (most of) the performance of the original larger model. Routing, as implemented in GPT-5, consists in directing a query, depending on its nature, to a right model. A complex problem would then be routed to a larger, more complex model while a basic calculation could directly be handled by a lightweight model.

> **Sustainable AI versus AI for sustainability**
>
> Sustainable AI targets the optimization of AI to makes it more environment friendly, e.g., by optimizing architecture and making the models more energy efficient. AI for sustainability considers the use of AI tools to achieve environmental benefits in other domains, e.g., optimizing energy grids to reduce CO2 emissions. Other names are often used such as green ICT (or just IT), ICT for green, AI for green, green AI…

**Optimization of electrical consumption.** Great engineering effort has been delivered, both by academia and companies, to make AI more sustainable and improve efficiency by orders of magnitude over short periods. This is a very active and fast-evolving area. Interestingly, DeepSeek has shown that a unique open-source initiative could demonstrate high performance at lower training consumption costs compared to closed-source counterparts [27]. Some later independent studies have nuanced the energy consumption gains for different reasons [8], [28]. Firstly, a lower cost may induce larger use (see rebound effect in Section 3.B). Secondly, the model can suffer from high inference costs, especially if the "chain-of-thought" reasoning gets activated, a technique that decomposes complex problems in multiple intermediate steps before giving the final result. More generally, reasoning is becoming more and more used by LLMs and may consume substantially more energy, outweighing efficiency gains.

Generally speaking, hardware and software optimization techniques can be broadly divided into three main areas. I) Model architecture and efficient algorithms. For instance, the use of quantization allows to store data and perform computations with less accuracy and thus cost. The Mixture-of-Experts (MoE) architecture only activates part of the model during inference (or training). II) Model deployment. For instance, using batch inference allows treating a maximum of queries in parallel to use all of the active computing power in a short amount of time and at a higher efficiency. III) Dedicated hardware and infrastructure: PUE optimization and use of dedicated AI-optimized hardware, going from central processing units (CPUs) to graphics processing units (GPUs) and tensor processing units (TPUs).

**Practical insights on the significance of the individual impact:** Finally, one could wonder about the significance of the individual use of AI. Figure 14 gives a comparison of the estimated electrical consumption of different typical Gen AI and daily life tasks. The estimations are based on sources and tools detailed in the caption. From observing it, we can conclude that a basic use of Gen AI consisting in writing a few emails per day, small conversations or summarizing an article has a moderate impact as compared to other daily tasks. Generation of images, long reports and especially videos have a larger impact, especially if they are repeated often.

The average per-capita Belgian electricity consumption is 132.17 Wh per 10 minutes, or equivalently about 19 kWh per day. From Figure 14, we can conclude that a basic daily usage consisting of generating a few emails, summarizing some articles and small conversations with chatbots would amount to a few tens of Wh per day, which is relatively small with respect to 19 kWh. If the goal of an individual is to significantly reduce their electricity consumption, this calls for a holistic vision across different sectors including, e.g., mobility. Reducing an AI basic usage would have a much lower impact than reducing the use of an electrical car and going for an (electric) bike.

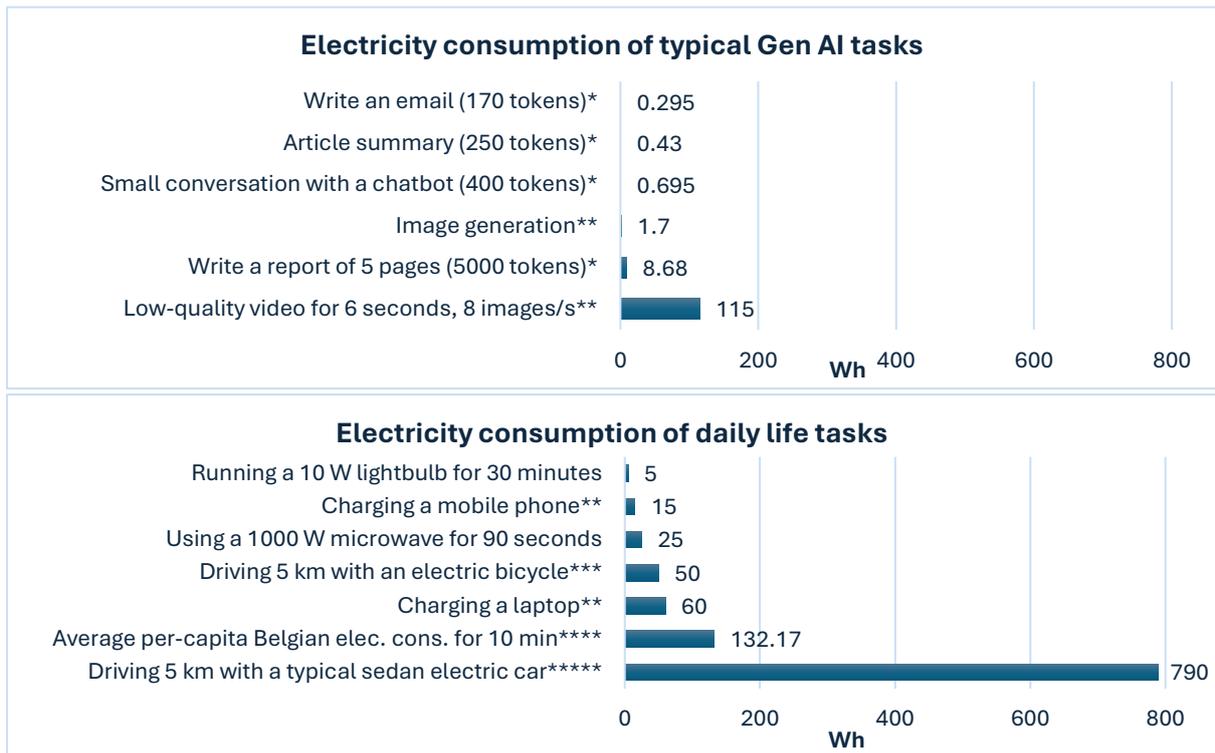

Figure 14 - Comparison of the estimated (utilization) electrical consumption of different Gen AI tasks with daily life tasks. *Using the EcoLogits Calculator with the GPT-5 model on Dec. 2025 (1 token ≈ 0.7 word). **Data from the IEA report [8]. ***Assuming that a typical 500 Wh battery has an autonomy of 50 km (can be smaller/larger depending on the case). ****The IEA website shows that the 2024 yearly electricity consumption per capita in Belgium was 6.947 MWh. *****Looking for a basic sedan car on Energy Watchers with reported consumption of 15.8 kWh/100km.

# 3. How to assess the direct and indirect impact of ICT and AI?

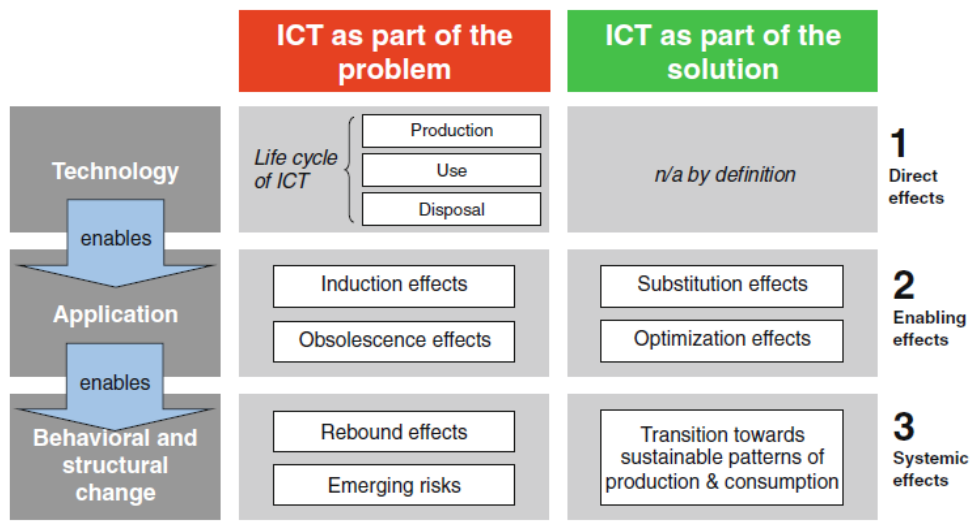

Figure 15 - Matrix to evaluate direct and indirect effects of ICT [29].

We have until now discussed mainly the direct (negative) impact of AI. What about the indirect positive and negative impacts? To help formalize this, L. M. Hilty and B. Aebischer have proposed the framework shown in Figure 15 treating ICT (or AI) as part of the problem or the solution [29].

This framework has been criticized (also by the authors) but has the merit to force the reader to look at the positive/negative and short/long term impacts. On a short term, direct effects are only negative for the environment as the ICT production, use and disposal lead to, e.g., water, resources and energy consumption. We discussed them in more detail in the previous section. On the other hand, indirect impact can be both positive and negative. The authors distinguish between shorter-term enabling effects and longer-term systemic effects.

### A. Positive indirect effects - enabled enhancements

As positive indirect effects, the use of AI for sustainability could reduce the use of another resource (optimization) or replace the use of another resource (substitution). It could also help the transition towards sustainable patterns of production and consumption. As an example, [Shayp](#) is a Belgian company using AI to detect water leakage. Another one comes from DeepMind AI stating that they could reduce their data center cooling bill thanks to… AI. The IEA report discusses in its 3$^{rd}$ chapter use cases of AI for energy optimization, with potential energy savings of up to 8% in light industry (manufacturing of electronics and machinery) and 20% for transport [8]. It can also help to reduce costs, which can be both beneficial or harmful for the environment. The IEA report emphasizes that oil and gas companies have been among the earlies adopters of AI technologies to boost exploration and exploitation. In this domain, AI could thus increase the production and affordability of fossil fuels, leading to increased emissions.

More generally, the EU Commission also sees digital services a key enabler to reach climate neutrality [4]. To justify these common claims, two often cited sources include a report from the GSMA association [30] and one from the GeSI [31]. They claim that a given carbon emission in the ICT domain would lead to a global net reduction up to 10 times larger, due to the enabled enhancements. Still, these studies have been criticized due to their speculative nature, the lack of evidence, the limited scientific rigor (often not peer-reviewed) and the fact that small-scope studies are used to extrapolate on a much larger scale. Generally speaking, even if specific AI use cases show clear environmental benefits, its overall positive balance is questionable, especially as a large share of AI use cases does not target an environmental benefit (AI for sustainability), e.g., creation of images or videos for entertainment, media sharing, marketing or recommendations.

### B. Negative indirect effects

As negative indirect effects, AI can stimulate the consumption of another resource (induction), shorten the useful life of another resource (obsolescence), leading to rebound effects, which seem particularly critical for AI. Rebound effect is a well-known economic effect and is often referred to as the Jevons' Paradox [32]. In the 19$^{th}$ century, the economist William Stanley Jevons observed that, as coal use became more efficient in steam engines, it paradoxically went together with an increase (not a decrease) of coal consumption. This effect can be explained by the economic principle of supply and demand. If a product becomes easier/cheaper to get, the price goes down and the demand can increase, creating a so-called "rebound effect". This effect could strongly affect the AI market, which is seeing drastic efficiency gains [33]. Let us take Google as an example. An article from the company [20] mentions a "33x reduction in energy consumption and a 44x reduction in carbon footprint for the median Gemini Apps text prompt over one year." Moreover, the Google 2024 environmental report [34] states that model optimization helped to "reduce the energy required to train an AI model by up to 100 times and reduce associated emissions by up to 1,000 times". They also improved AI hardware efficiency: "For example, our TPU v4 was 2.7 times more energy efficient than our TPU v3,54 and we'll soon offer Nvidia's next-generation Blackwell GPU to Cloud customers, which Nvidia estimates will train large models using 75% less power than older GPUs to complete the same task." All of this means formidable engineering effort and technological progress. Still, despite these massive efficiency gains, the company stated itself that the further integration of AI may challenge reducing emissions due to the increasing energy demands [1].

In Figures 1 – 3, we can also observe that the impact has worsened over the last years. These figures do not only account for AI but also data centers more generally. Related numbers, i.e., global trends in digital and energy indicators, are given in the table of Figure 16 for the period 2015-2022 [11]. While the data center workload has increased by 340%, energy efficiency improvements have helped to limit the energy consumption growth to 20-70%. In other words, the same task requires now less energy. Still, even if the efficiency has greatly increased over time, the footprint of the utilization (number of users, queries, size of the models, use of reasoning, infrastructure...) has increased even faster, resulting in larger emissions in the end.

Efficiency improvement is the typical sustainability objective in ICT and AI. This should be addressed carefully as it might create a rebound effect and stimulate more consumption, which in the end could worsen the situation.

|  | 2015 | 2022 | Change |
| --- | --- | --- | --- |
| Internet users | 3 billion | 5.3 billion | +78% |
| Internet traffic | 0.6 ZB | 4.4 ZB | +600% |
| Data centre workloads | 180 million | 800 million | +340% |
| Data centre energy use (excluding crypto) | 200 TWh | 240–340 TWh | +20–70% |
| Crypto mining energy use | 4 TWh | 100–150 TWh | +2300–3500% |
| Mobile subscriptions | 7.1 billion | 8.3 billion | +17% |
| Fixed broadband subscriptions | 790 million | 1.2 billion | +51% |
| Data transmission network energy use | 220 TWh | 260–360 TWh | +18–64% |

Sources: IEA (2023a); Malmodin et al. (2023)

Figure 16 - Global trends in digital and energy indicators, 2015–2022 [11].

## 4. What can we do?

We here give a list of practical tips, individual and collective actions based on the previous discussions, which can be used to optimize AI impact. This section has a looser scientific foundation than previous sections. Yet, it intends to give directions to people who asked us how they could work with AI responsibly. Coming to actions, even based on scientific data, requires making choices. As shown in Figure 17, we cluster these actions around 3 themes: understand, reflect and act responsibly.

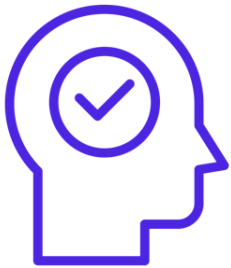 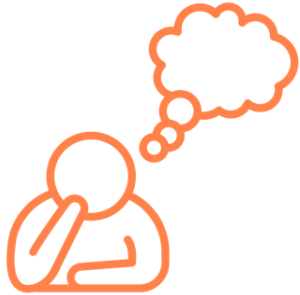 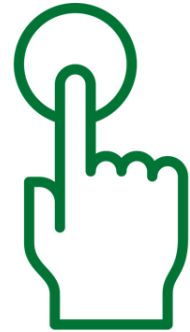

Understand          Reflect          Act responsibly

Figure 17 – Three clusters to structure positive actions.

Some tips below refer to individual measures. The goal is not to make individuals feel guilty or to shift the blame. In that regard, the inaction triangle, shown in Figure 18 and theorized by Pierre Peyretou, can provide an interesting model to understand the challenges of the situation and potential solutions. This model illustrates the problem of shifting the burden onto other actors. Applied to our case, citizens could blame big tech AI companies for not having sufficient climate ambitions and question their personal responsibility, especially as AI use becomes invisible and automatized, e.g., just by visiting a website or using an app. AI companies could justify themselves by replying that they only respond to customers' and stakeholders' needs. Politicians could argue that they align with the citizens' wishes. In short, each one has a good reason not to do take responsibility. To break this vicious triangle (circle?) and turn it into a virtuous one, it is important to understand, to reflect, and then to develop active collaborations between all actors to act simultaneously. This resonates with the three clusters of Figure 17.

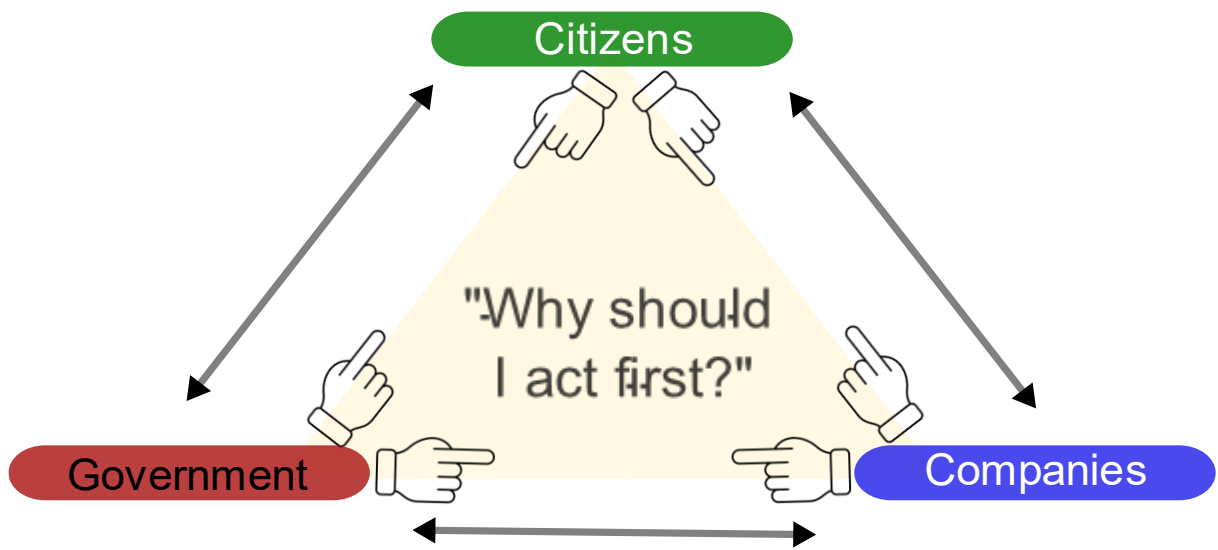

Figure 18 - Inaction triangle (credit : Pierre Peyretou).

**Understand**

- <u>Increase AI literacy</u>. We should improve our knowledge about AI, and we can make that fun. There exist for instance many board and collective games, collective workshops around AI, e.g. The Digital Collage https://digitalcollage.org/; The AI® Mural https://lafresquedelia.com/. A number of online websites, such as, e.g., [22], [25], provide interesting information to further investigate and form an opinion based on different sources.
- Use tools to measure and reduce your footprint. Here are a few interesting tools:
    - EcoLogits calculator https://huggingface.co/spaces/genai-impact/ecologits-calculator providing consumption of generative AI models for different vendors and different uses, including embodied emissions, multiple criteria and clear comparisons. The methodology is open and transparent to the user.
    - Tips towards Green IT https://altimpact.fr/en/
    - Code Carbon: "Track and reduce CO2 emissions from your computing", "CodeCarbon is a lightweight software package that seamlessly integrates into your Python codebase." https://codecarbon.io/
- Raise awareness. Discuss it around you in your personal and professional contexts, share useful information and tools.

**Reflect**

- Ask for more transparency and accountability: get mobilized to ask for more transparency from your AI providers. When you buy a car, you know its consumption for 100 km, why not the same transparency for AI tools? Sasha Luccioni, an AI and climate researcher at Hugging Face, has created the AI Energy Score, an energy efficiency metric for AI models. It should become mandatory for companies to provide some such metrics. The EU AI Act, which entered into force from the 1st of August 2024 and with the majority of the provisions on August 2026, represents a significant move in this direction.
- Question the dream: beyond environmental concerns, explore further: which world/society do you want? Which part do you want AI to play? What power/role should AI have? In relation, (re-)watch the movie WALL-E.
- Rethink about AI in the light of the 6 Rs of a sustainable circular economy : refuse, rethink, reduce, reuse, repurpose, recycle. This is particularly important as AI use is often pushed on people. More and more, the use of AI and LLMs become invisible. By choosing the software, the hardware, the applications, we can have an impact on how we use AI and its environmental impact.

**Act responsibly**

- Use your own HI (Human Intelligence): think before you prompt. The most impactful phase of the development and deployment of AI is the inference phase. We can reduce our individual and collective use of generative AI and only use it when needed. In the light of Figure 14, this reflection should especially be conducted
- Consider sufficiency and efficiency in line with the sufficiency, efficiency, renewables (SER) framework put forward by the IPCC 6th assessment report:
    - Sufficiency: question the need of using AI. Do I need it? What is the benefit? What are the costs and benefits related to the environmental, economic and social dimensions? Is there a net benefit?
    - Efficiency: if you find the need relevant, use the right tool, the more environmental friendly. Other criteria can/should play a role in the decision such as going for a European AI tool (e.g. Mistral AI, DeepL), ethical, open source, more transparent on its environmental impact…
- Model size matters: use small models. Avoid generalist AI tools. If an AI needs to be multifunctional and be able to perform a large variety of tasks, it requires much more training, data and model complexity. Use dedicated tools specialized for your use.

Example: AI tool for translating text (e.g. DeepL, Google Translate). As previously discussed, as routing is becoming more implemented by LLM providers, a query could become automatically routed to a smaller model as a function of its nature.
- <u>Prompt and output size matters</u>: use shorter prompts. Keep your messages concise and to the point of reducing token consumption (Less tokens = less model calls = less consumption). To get a shorter output, append "Be Concise" to your prompt. This last aspect might be less useful as/when reasoning models are used. Even if the output is concise, it might have generated a lot of text to arrive to it.

Figure 19 in the appendix also provides a visual illustration (with text in Dutch) of practical tips.

We here looked at a number of tips on the environmental impact of Gen AI. While this is important, if our objective is to limit our environmental impact, we should also embrace a more holistic vision and look broader as other domains might have a strong impact.

## 5. Closing remarks

We live in the data era and AI has shown (is showing) great capabilities, yet comes at an ecological cost. This cost is hard to exactly assess quantitatively. Yet its substantial and rising nature is undeniable.

This calls for action. It is high time to raise awareness, increase literacy, progress towards responsible usage, and finally reflect on the place we want for AI in our life.

The R&D community is progressing in this direction, proposing methods and more efficient solutions, including approaches to better compare and assess direct/indirect impacts. In that regard, the authors of this report engage in designing and applying a sustainability assessment framework in the context of the EU-funded Horizon Europe SUSTAIN-6G project.

## 6. References

[1] Google, "Environmental Report 2025." [Online]. Available: https://sustainability.google/reports/google-2025-environmental-report/
[2] Microsoft, "2025 Microsoft Environmental Sustainability Report." [Online]. Available: https://www.microsoft.com/en-us/corporate-responsibility/sustainability/report/
[3] I. O'Brien, "Data center emissions probably 662% higher than big tech claims. Can it keep up the ruse?," *The Guardian*, Sept. 15, 2024. Accessed: Oct. 15, 2025. [Online]. Available: https://www.theguardian.com/technology/2024/sep/15/data-center-gas-emissions-tech
[4] "COMMUNICATION FROM THE COMMISSION The European Green Deal." Dec. 2019. Accessed: Oct. 21, 2025. [Online]. Available: https://eur-lex.europa.eu/legal-content/EN/TXT/?uri=CELEX:52019DC0640
[5] European Commission. Joint Research Centre., *Understanding Product Environmental Footprint and Organisation Environmental Footprint methods*. LU: Publications Office, 2022. Accessed: Sept. 23, 2025. [Online]. Available: https://data.europa.eu/doi/10.2760/11564
[6] C. Freitag, M. Berners-Lee, K. Widdicks, B. Knowles, G. S. Blair, and A. Friday, "The real climate and transformative impact of ICT: A critique of estimates, trends, and regulations," *Patterns*, vol. 2, no. 9, p. 100340, Sept. 2021, doi: 10.1016/j.patter.2021.100340.
[7] "Aviation," IEA. Accessed: Sept. 23, 2025. [Online]. Available: https://www.iea.org/energy-system/transport/aviation
[8] IEA, "Energy and AI – Analysis," Apr. 2025. Accessed: Aug. 14, 2025. [Online]. Available: https://www.iea.org/reports/energy-and-ai

[9]  Fluvius, "System operators Elia and Fluvius upgrade action plans for congestion management for businesses." Accessed: Dec. 19, 2025. [Online]. Available: https://pers.fluvius.be/system-operators-elia-and-fluvius-upgrade-action-plans-for-congestion-management-for-businesses

[10] "Raccordement électrique en Wallonie : pourquoi l'accès à la puissance devient un défi ?" Accessed: Dec. 19, 2025. [Online]. Available: https://www.ores.be/professionnel/acces-a-la-puissance

[11] European Commission. Joint Research Centre., "Energy consumption in data centres and broadband communication networks in the EU.," Publications Office, LU, 2024. Accessed: June 11, 2024. [Online]. Available: https://data.europa.eu/doi/10.2760/706491

[12] S. J. Russell and P. Norvig, *Artificial intelligence: a modern approach*, Fourth Edition. in Pearson Series in Artificial Intelligence. Hoboken, NJ: Pearson, 2021.

[13] C.-J. Wu *et al.*, "Sustainable AI: Environmental Implications, Challenges and Opportunities," in *Proceedings of Machine Learning and Systems*, arXiv, Jan. 2022. doi: 10.48550/arXiv.2111.00364.

[14] P. Villalobos, A. Ho, J. Sevilla, T. Besiroglu, L. Heim, and M. Hobbhahn, "Position: Will we run out of data? Limits of LLM scaling based on human-generated data," in *Proceedings of the 41st International Conference on Machine Learning*, in ICML'24. Vienna, Austria: JMLR.org, 2024, p. 22.

[15] E. Strubell, A. Ganesh, and A. McCallum, "Energy and Policy Considerations for Modern Deep Learning Research," *Proc. AAAI Conf. Artif. Intell.*, vol. 34, no. 09, pp. 13693–13696, Apr. 2020, doi: 10.1609/aaai.v34i09.7123.

[16] A. S. Luccioni, S. Viguier, and A.-L. Ligozat, "Estimating the Carbon Footprint of BLOOM, a 176B Parameter Language Model," *J. Mach. Learn. Res.*, vol. 24, no. 253, pp. 1–15, 2023.

[17] P. Jiang, C. Sonne, W. Li, F. You, and S. You, "Preventing the Immense Increase in the Life-Cycle Energy and Carbon Footprints of LLM-Powered Intelligent Chatbots," *Engineering*, vol. 40, pp. 202–210, Sept. 2024, doi: 10.1016/j.eng.2024.04.002.

[18] "Carbon footprint of travel per kilometer," Our World in Data. Accessed: Dec. 11, 2025. [Online]. Available: https://ourworldindata.org/grapher/carbon-footprint-travel-mode

[19] "$CO_2$ emissions per capita," Our World in Data. Accessed: Dec. 11, 2025. [Online]. Available: https://ourworldindata.org/grapher/co-emissions-per-capita

[20] C. Elsworth *et al.*, "Measuring the environmental impact of delivering AI at Google Scale," Aug. 21, 2025, *arXiv*: arXiv:2508.15734. doi: 10.48550/arXiv.2508.15734.

[21] S. Samsi *et al.*, "From Words to Watts: Benchmarking the Energy Costs of Large Language Model Inference," in *2023 IEEE High Performance Extreme Computing Conference (HPEC)*, Boston, MA, USA: IEEE, Sept. 2023, pp. 1–9. doi: 10.1109/HPEC58863.2023.10363447.

[22] J. You, "How much energy does ChatGPT use?," Epoch AI. Accessed: Dec. 14, 2025. [Online]. Available: https://epoch.ai/gradient-updates/how-much-energy-does-chatgpt-use

[23] A. Chatterji *et al.*, "How People Use ChatGPT," National Bureau of Economic Research, Cambridge, MA, w34255, Sept. 2025. doi: 10.3386/w34255.

[24] S. Luccioni, Y. Jernite, and E. Strubell, "Power Hungry Processing: Watts Driving the Cost of AI Deployment?," in *Proceedings of the 2024 ACM Conference on Fairness, Accountability, and Transparency*, in FAccT '24. New York, NY, USA: Association for Computing Machinery, June 2024, pp. 85–99. doi: 10.1145/3630106.3658542.

[25] J. O'Donnella and C. Crownhart, "We did the math on AI's energy footprint. Here's the story you haven't heard.," MIT Technology Review. Accessed: Dec. 17, 2025. [Online]. Available: https://www.technologyreview.com/2025/05/20/1116327/ai-energy-usage-climate-footprint-big-tech/

[26] J. Gerstmayr, P. Manzl, and M. Pieber, "Multibody Models Generated from Natural Language," *Multibody Syst. Dyn.*, vol. 62, no. 2, pp. 249–271, Oct. 2024, doi: 10.1007/s11044-023-09962-0.

[27] Z. Deng *et al.*, "Exploring DeepSeek: A Survey on Advances, Applications, Challenges and Future Directions," *IEEECAA J. Autom. Sin.*, vol. 12, no. 5, pp. 872–893, May 2025, doi: 10.1109/JAS.2025.125498.

[28] J. O'Donnell, "DeepSeek might not be such good news for energy after all," MIT Technology Review, Jan. 2025. Accessed: Oct. 27, 2025. [Online]. Available: https://www.technologyreview.com/2025/01/31/1110776/deepseek-might-not-be-such-good-news-for-energy-after-all/

[29] L. M. Hilty and B. Aebischer, Eds., *ICT Innovations for Sustainability*, vol. 310. in Advances in Intelligent Systems and Computing, vol. 310. Cham: Springer International Publishing, 2015. doi: 10.1007/978-3-319-09228-7.

[30] GSMA, "The Enablement Effect," 2019. [Online]. Available: https://www.gsma.com/solutions-and-impact/connectivity-for-good/external-affairs/gsma_resources/the-enablement-effect/

[31] "#SMARTer2030 ICT Solutions for 21st Century Challenges," 2015. [Online]. Available: https://smarter2030.gesi.org/

[32] W. S. Jevons, "The Coal Question," *Econ. Popul. Routledge*, 204 193 AD.

[33] A. S. Luccioni, E. Strubell, and K. Crawford, "From Efficiency Gains to Rebound Effects: The Problem of Jevons' Paradox in AI's Polarized Environmental Debate," in *Proceedings of the 2025 ACM Conference on Fairness, Accountability, and Transparency*, Athens Greece: ACM, June 2025, pp. 76–88. doi: 10.1145/3715275.3732007.

[34] Google, "Environmental Report 2024." [Online]. Available: https://sustainability.google/reports/google-2024-environmental-report/

# 7. Annex

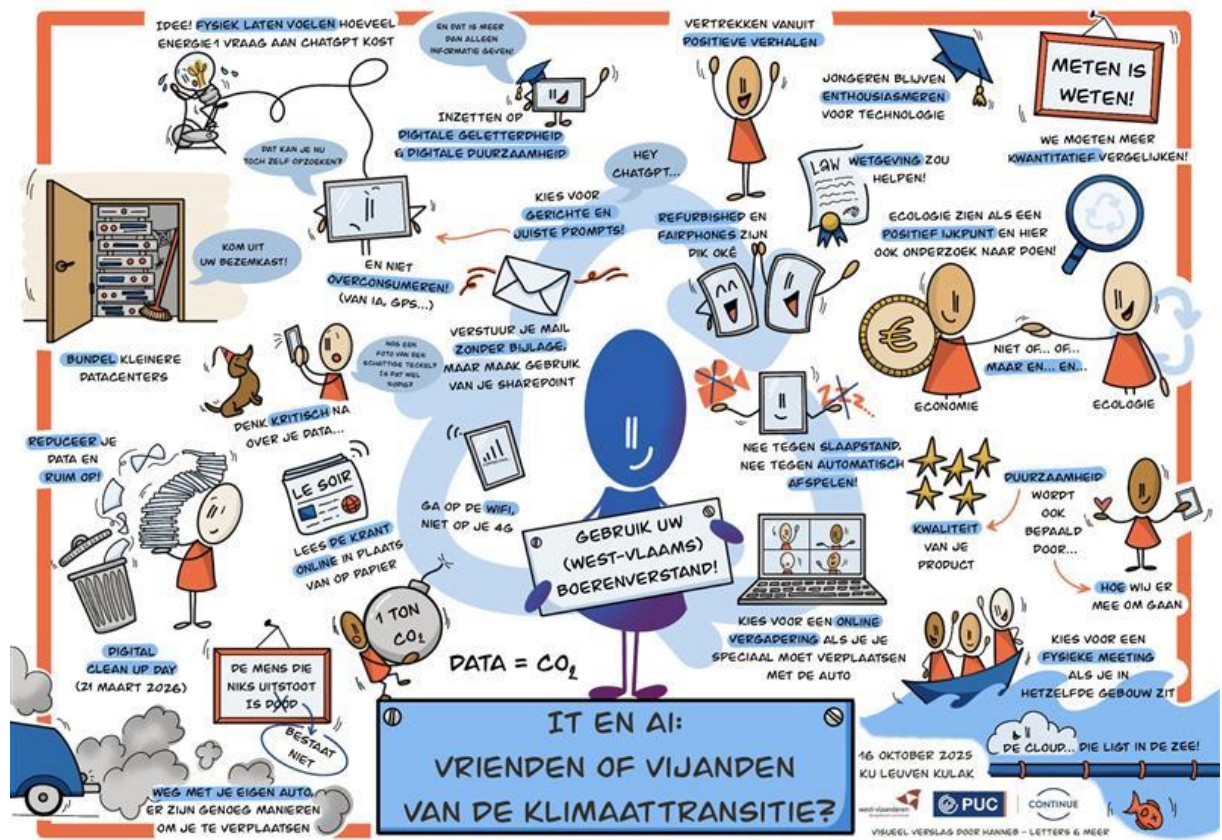

Figure 19 - Visual from IT/AI study evening in Kortrijk.